%% file: main.tex
\documentclass[letterpaper, 10 pt, conference]{ieeeconf}
\IEEEoverridecommandlockouts

\usepackage{graphics} 
\usepackage{epsfig} 
\usepackage{times} 

\usepackage{amsthm}
\usepackage{amsmath} 
\usepackage{amssymb}  
\usepackage{mathtools}
\usepackage[short]{optidef}
\usepackage{tensor}
\usepackage{units}
\usepackage{tikz}
\usetikzlibrary{decorations.pathreplacing,calc}
\usepackage{pgfplots}
\pgfplotsset{compat=newest}
\usepackage{url}

\usepackage{relsize}
\usepackage{booktabs}
\usepackage{tikz}
\usetikzlibrary{calc,patterns,angles,quotes}
\usepackage{colortbl}%
\newcommand{\trowgray}{\rowcolor[gray]{0.925}}
\usepackage{eqparbox}

\definecolor{CCPSgray3}{RGB}{242,242,242}
\definecolor{CCPSgreen1}{RGB}{67,80,58}
\definecolor{mlred}{RGB}{208,2,27}
\definecolor{MPIgreen}{RGB}{0,118,117}

\usepackage{algorithmic}
\usepackage[inkscapeformat=png]{svg}
\usepackage{graphicx}
\usepackage{textcomp}
\usepackage{xcolor}
\def\BibTeX{{\rm B\kern-.05em{\sc i\kern-.025em b}\kern-.08em
    T\kern-.1667em\lower.7ex\hbox{E}\kern-.125emX}}
    
\theoremstyle{plain}
\newtheorem{assumption}{Assumption}

\theoremstyle{plain}
\newtheorem{corollary}{Corollary}

\begin{document}
\bstctlcite{IEEEexample:BSTcontrol}

\title{\LARGE \bf Neural Horizon Model Predictive Control - Increasing Computational Efficiency with Neural Networks}
\author{H. Alsmeier$^{1}$, A. Savchenko$^{1}$, and R. Findeisen$^{1}$
 \thanks{$^{1}$  Control and Cyber-Physical Systems Laboratory, Technical University of Darmstadt, Germany,\newline \{hendrik.alsmeier, anton.savchenko, rolf.findeisen\}@iat.tu-darmstadt.de}}
\maketitle
\thispagestyle{empty}
\pagestyle{empty}


\begin{abstract}
The expansion in automation of increasingly fast applications and low-power edge devices poses a particular challenge for optimization based control algorithms, like model predictive control. Our proposed machine-learning supported approach addresses this by utilizing a feed-forward neural network to reduce the computation load of the online-optimization. We propose approximating part of the problem horizon, while maintaining safety guarantees -- constraint satisfaction -- via the remaining optimization part of the controller. The approach is validated in simulation, demonstrating an improvement in computational efficiency, while maintaining guarantees and near-optimal performance. The proposed MPC scheme can be applied to a wide range of applications, including those requiring a rapid control response, such as robotics and embedded applications with limited computational resources.
\end{abstract}





\vspace{-.5em}\section{Introduction}
\label{sec:into}

From its roots in the process industry, model predictive control (MPC) has been established as a safe and reliable control algorithm for a wide range of systems, from autonomous driving and robotics to (bio)chemical plants. Its wide adoption has been driven by MPCs flexibility of formulating control problems, satisfying constraints, and enabling optimal operation \cite{rawlings2017,findeisen2002}. However, MPC can lose real-time feasibility in case the plant evolves faster than the algorithm can solve the underlying optimal control problem (OCP)~\cite{rawlings2017,zometa2012implementation}, e.g. if the system model consists of many states, or its prediction horizon is too long. Thus, the speedup of the OCP is a popular research topic for MPCs~\cite{diehl2009efficient}.

One method to accelerate computations is to solve the MPC problem to sub-optimality, as demonstrated, for example, by the real-time iteration approach outlined in~\cite{diehl2002real}, which linearizes the OCP to solve a quadratic problem.

Another approach is explicit MPC, which aims to determine an analytic pre-computed MPC solution. It is, however, typically limited to linear systems with quadratic cost and small state dimensions, as the solution consists of numerous linearly affine control laws \cite{Zeilinger2011}.

 Machine learning (ML) can be used to approximate the implicit MPC control law and replace the controller in the closed-loop with an approximation. In~\cite{Borrelli2019} the implicit control law of a MPC controller was learned from data samples of measured states and resulting MPC control actions. By avoiding online optimization such imitation controllers reduce computational burden but also lose explicit contraint satisfaction and optimality.

To utilize advantages of both ML and traditional optimization-based predictive control, learning-supported MPC algorithms have started developing. Such approaches, for example, fuse a ML model with a MPC by replacing parts of the controller with a learned mapping, while retaining the optimization to ensure safety. The method presented in this paper falls into this category, as we seek to replace part of the horizon and cost function in the OCP.
Firstly, the entire model as well as unknown or complex effects of it can be learned. Approaches of this kind either aim to reduce the model complexity \cite{Lucia2020,Lanzetti2019} or increase model precision~\cite{Salzmann2023,Wu2023,Borrelli2018}. All such approaches approximate state transition mapping from collected data, which is also the foundation of our method.
Secondly, an objective function of the OCP can be learned, e.~g. to make scenario-based approaches computationally tractable \cite{Lucia2021,Lucia2021b}, or to convexify the OCP \cite{Gravdahl2022}.
In some cases, only the terminal ingredients are learned, e.~g. to shorten the horizon and learn a convex terminal cost \cite{Bemporad21}, or even approximate the infinite horizon cost \cite{Streif2022}.

Our contribution combines these aspects of model and cost function learning. We approximate the open-loop optimal state sequence, which we then use to construct a learned terminal cost for a shorter MPC horizon. We show that doing so leads to more stable results than learning the optimal cost function directly, while ensuring recursive feasibility under suitable conditions.
The paper is structured as follows: Section~\ref{sec:problemsetup} motivates our approach and provides background on MPC and neural networks. Section \ref{sec:neuralhorizon} describes the Neural Horizon MPC approach and establishes recursive feasibility results. In Section \ref{sec:simresults} we validate the approach in simulation before concluding with final remarks.

\section{Problem Setup}
\label{sec:problemsetup}
We consider a discretized continuous-time nonlinear system:\vspace{-0.75em}
\begin{equation}
        x(t_i+1) = f(x(t_i),u(t_i)). \label{eq:dynamics1}
\end{equation}
Here $x(t_i) \in \mathbb{R}^{n_x}$ represents the system states, $u(t_i) \in \mathbb{R}^{n_u}$ are the applied inputs for the current time point $t_i$. The function $f: \mathbb{R}^{n_x} \times \mathbb{R}^{n_u} \rightarrow \mathbb{R}^{n_x}$ denotes the state transition mapping.\\
We consider a model predictive controller (MPC), which works as follows~\cite{rawlings2017,findeisen2002}. At each sampling point $t_i$ it solves an optimal control problem (OCP) and applies its first optimal input to the system \eqref{eq:dynamics1} until the next sampling time $t_{i+1}$.\\
By repeating this process from the next sampling point $t_{i+1}$ we close the control loop, applying the optimal input at every timestep~\cite{rawlings2017,findeisen2002}. The underlying OCP for model dynamics \eqref{eq:dynamics1} can be summarized in the form of a nonlinear optimization program (NLP):\\
\begin{argmini}[1]
    {\substack{\{\!x_k\!\}_0^N\!,\,\{\!u_k\!\}_0^{N\!-\!1}}}{\hspace{-.5cm}L\left(\{\!x_k\!\}_0^{N-1}\!,\{\!u_k\!\}_0^{N-1}\right)\!+\!V(x_N)}{ \label{eq:mpc}}{\mathrm{OCP}(x_{\mathrm{init}})=\hspace{-.35cm}}
    \addConstraint{x_{k+1}}{=f\left(x_k,u_k\right)}{\ \forall k\in [0,\hdots,N\!-\!1]}
    \addConstraint{x_k}{\in\mathcal{X}_k}{\ \forall k\in [0,\hdots,N]}
    \addConstraint{u_k}{\in\mathcal{U}_k}{\ \forall k\in [0,\hdots,N\!-\!1]}
    \addConstraint{x_0}{=x_{\mathrm{init}}}.
\end{argmini}
We denote with $k$ the discrete time index within the OCP, and with $N$ the full horizon length.
Sets $\{\!\xi_k\!\}_a^b:=\{\xi_a,\xi_{a+1},\hdots,\xi_b\}$ denote the time series of the corresponding vectors. Thus, $\{\!x_k\!\}_0^N$ and $\{\!u_k\!\}_0^{N-1}$ are the time series of decision variables for states and inputs correspondingly. $x_{\text{init}}$ is the state of the system at the sampling time. The sets $\mathcal{X}_k$ and $\mathcal{U}_k$ denote generalized state and input constraints, as well as the terminal set $\mathcal{X}_N$, which we assume to be compact and containing the origin. The cost function is comprised of two parts -- the sum of \emph{stage costs} for each time point $k\in[0,\hdots,N-1]$, that we denote as $L\left(\{\!x_k\!\}_0^{N-1}\!,\{\!u_k\!\}_0^{N-1}\right)$, and the \emph{terminal cost} $V(x_N)$.\\
In the terms of the OCP \eqref{eq:mpc}, the MPC feedback is given by:
\begin{equation}
    u^{*}(t_i) = \mathrm{MPC}\left(x(t_i)\right) := \mathrm{OCP}\left(x(t_i)\right)|_{u_0} \label{eq:MPCmapping}.
\end{equation}
The problem \eqref{eq:mpc} can be solved in different ways, e.g. via iterative numerical optimization. The computational time of one iteration is in general high and depends on the structure of the OCP, the horizon
length, and the system size. We employ the observation, that shortening horizon length $N$ generally should reduce computational complexity.

\subsection{Neural Networks}
\label{subsec:NNs}

We chose neural networks (NNs) to approximate the needed mappings, as they fulfill the universal approximation theorem~\cite{hornik1991,hornik1993,scarselli1998} and can therefore approximate any mapping to an arbitrary precision. The simplest NNs consist of neurons organized into layers, where the value of each layer's neurons is computed as an affine transformation of the neurons at a previous layer, passed through a nonlinear activation function~\cite{ojha2017}. For the vector of neurons $z_j$ of the layer $j$ we can represent this in the matrix form:\vspace{-1 ex}
\begin{equation}
    z_j = \alpha_j\circ\zeta_j(z_{j-1}):= \alpha_j \left( W_{j} z_{j-1} + b_j \right),
    \label{eq:nnlayer}
\end{equation}
where $\alpha_j$ represents a element-wise activation function, $z_{j-1}$ denotes the vector of the previous layer's neurons and $W_{j}$, $b_j$ are called the \emph{weights} and \emph{biases} of layer $j$. The layer $0$ is denoted as the input layer, and The last layer is called the output layer. The remaining layers are called hidden layers, and the NN is called \emph{deep} if it has more than one hidden layer. The NN with $m-1$ hidden layers, transforming input $x$ to output $y$, can be written as follows:\vspace{-1 ex}
\begin{equation}
 y = NN(x):= \zeta_m \circ \alpha_{m-1}\circ \zeta_{m-1} \circ \hdots \circ \alpha_1 \circ \zeta_1(x) 
\label{eq:nndesign}
\end{equation}
We employ fully connected layers (i.~e. all weight matrices $W_j$ are full) with hyperbolic tangent activation function for all hidden layers.


\section{Neural Horizon MPC}
\label{sec:neuralhorizon}

We propose simplifying the OCP formulation \eqref{eq:mpc} over $k \in [M+1,\hdots,N]$ for some $M \ll N$. This procedure removes decision variables by replacing the state predictions with a computationally cheap approximated NN-based model. We denote these approximated sequences of states as $ \{\! \tilde x_k\!\}_{M+1}^{N}$, and the corresponding mapping as\vspace{-1 ex}
\begin{equation}
    \{\! \tilde x_k\!\}_{M+1}^{N} = \tilde f_{s}(x_M).\label{eq:nn_state_model}
\end{equation}
The function $\tilde f_{s}(\cdot)$ is a neural network of the form \eqref{eq:nndesign}, generating an approximated optimal open-loop state sequence $ \{\! \tilde x_k\!\}_{M+1}^{N}$ based on the state $x_M$, which remains a decision variable. $ \{\! \tilde x_k\!\}_{M+1}^{N}$ replaces the tail of the optimal solution trajectory of the OCP \eqref{eq:mpc}. This sequence is then employed to calculate the cost function over the horizon $k \in [M+1,\hdots,N]$, approximating the cost-to-go with optimal $u_k$ of the OCP \eqref{eq:mpc} over the horizon $M$ to $N$. This approach removes dependencies of the states on inputs $\{\! u_k\!\}_{M+1}^{N-1}$, since we outright generate the optimal state sequence given this optimal input sequence. And as the neural network produces the entire sequence at once, fewer function evaluations of $\tilde f_{s}(\cdot)$ are needed during optimization. The modified OCP, that we denote \emph{Neural Horizon MPC}, we thus formulate as
\begin{argmini}[2]
    {\substack{\{\!x_k\!\}_0^M\!,\, \{\!u_k\!\}_0^{M}}}{\hspace{-.2cm}L \left(\{\!x_k\!\}_0^{M},\{\!u_k\!\}_0^{M} \right)\!+\!\tilde L \left(\{\!\tilde x_k\!\}_{M\!+\!1}^{N\!-\!1} \right)\!+\!\tilde V\left(\tilde{x}_N \right)}{\label{eq:neural_MPC}}{}
    \addConstraint{x_0}{=x_{\mathrm{init}}}
    \addConstraint{x_{k+1}}{= f\left(x_k,u_k\right)}{\forall k \in [0,\hdots,M\!-\!1]}
    \addConstraint{x_k}{\in\mathcal{X}_k}{\forall k \in [0,\hdots,M]}
    \addConstraint{u_k}{\in\mathcal{U}_k}{\forall k \in [0,\hdots,M]}
    \addConstraint{\{\! \tilde x_k\!\}_{M+1}^{N}}{=\tilde{f}_{s} \left(x_M \right)}{}
    \addConstraint{\tilde{x}_k}{\in \mathcal{\tilde{X}}_k}{\forall k \in [M\!+\!1,\hdots,N]}
\end{argmini}
In contrast to \eqref{eq:mpc}, current formulation introduces the approximated stage costs $\tilde L(\{\!\tilde x_k\!\}_{M\!+\!1}^{N\!-\!1})$, terminal cost $\tilde V_f(\tilde x_N)$, state sequence $\{\! \tilde x_k\!\}_{M+1}^{N}=\tilde{f}_{s}(x_M)$, and state constraints $\tilde{\mathcal{X}_k}$.
Since the state sequence is generated from a static mapping $\tilde f_{s}(x_M)$, it can be viewed as a more elaborate terminal cost and terminal constraint for the OCP \eqref{eq:mpc} with a horizon $M$. For process safety, it is to note that we retain the ability to enforce constraints over the Neural Horizon $x_k \in \tilde{\mathcal{X}}_k$. Furthermore, introduced cost functions can be separately tuned to compensate for possible state prediction errors.

In the simplest case, we can reuse all components of \eqref{eq:mpc} for the Neural Horizon MPC, such as state constraints $\tilde{\mathcal{X}_k} = \mathcal{X}_k$ and the corresponding components of the cost function. However, as will be shown in Section~\ref{subsec:stab}, some adjustments are needed for stability guarantees.

\subsection{Cost Estimation of Neural Horizon}
\label{subsec:nhconstruction}
The mapping \eqref{eq:nn_state_model} converts a state of dimension $n_x$ to a state sequence of length $N\!-\!M$, i.e. $\tilde f_{s}:\mathbb R^{n_x}\to \mathbb R^{n_x\times (N\!-\!M)}$.
Alternatively, we can consider approximating the cost function directly, mapping the state $x_M$ to the value function over $k \in [M\!+\!1,\hdots,N]$ as follows\vspace{-1 ex}
\begin{equation}
     \tilde f_c \left( x_M \right)\approx\begin{bmatrix}  L \left(\{\! x_k\!\}_{M\!+\!1}^{N\!-\!1} \right)\!+\! V\left(x_N \right), &  L \left(\{\!u_k\!\}_{M\!+\!1}^{N\!-\!1} \right)\end{bmatrix}.\label{eq:nn_cost_model}
\end{equation}
Both components generated by $\tilde f_c \left( x_M \right)$ can then be added directly to the shortened cost function of the OCP, analog to \eqref{eq:neural_MPC}. In this formulation the state predictions are not made and thus cannot be explicitly bounded.

Even though such a formulation should be equivalent to \eqref{eq:neural_MPC} without constraints on $\tilde{x}_k$ (given trained NNs of similar quality), the simulation results show a large discrepancy between the approaches, cf. Section~\ref{sec:simresults} for more details.

\subsection{Recursive Feasibility and Constraint Satisfaction}
\label{subsec:stab}

To assess the recursive feasibility and constraint satisfaction of the presented Neural Horizon MPC design we introduce the following Assumption:
\begin{assumption}\label{as:1}
    Given a continuously differentiable mapping $\Phi\colon\mathcal A\to \mathcal B\colon a\mapsto b$ and an error bound $\gamma >0$ there exists a representative dataset $\{(a_i,b_i)\}_{i\in \mathcal I}$ and a training procedure for the neural network of type \eqref{eq:nndesign}, which results in $NN(a)\in \mathcal B$ and $\lVert NN(a)-\Phi(a)\rVert_{\infty}<\gamma$ for all $a\in \mathcal A$.
\end{assumption}
As the neural networks are general approximators, in our view it is reasonable to expect the properties in Assumption~\ref{as:1} to hold for sufficiently large networks and training sets, even if guaranteeing it is difficult.

This assumption can now be used to generate the training data in a way, that the predictions are guaranteed to satisfy given state constraints $\mathcal X$. For that, we adapt the OCP \eqref{eq:mpc} as follows:\vspace{-0.5em}
\begin{argmini*}[1]
    {\substack{\{\!x_k\!\}_0^N\!,\,\{\!u_k\!\}_0^{N\!-\!1}}}{L\left(\{\!x_k\!\}_0^{N-1}\!,\{\!u_k\!\}_0^{N-1}\right)\!+\!V(x_N) }{\label{eq:mpc_tightend}}{\Phi_e(x_{\mathrm{init}})=\!}
    \addConstraint{x_{k+1}}{=f\left(x_k,u_k\right)}{\ \forall k\in [0,\hdots,N\!-\!1]}
    \addConstraint{x_k}{\in\mathcal{X}}{\ \forall k\in [0,\hdots,M]}
    \addConstraint{x_k}{\in\mathcal{X} \ominus \mathbf{B}_{e}}{\ \forall k\in [M+1,\hdots,N]}
    \addConstraint{u_k}{\in\mathcal{U}_k}{\ \forall k\in [0,\hdots,N\!-\!1]}
    \addConstraint{x_0}{=x_{\mathrm{init}}}.
\end{argmini*}
Here, $\mathbf{B}_{e}$ denotes the ball of size $e$ for the infinity norm, and $\ominus$ is a Pontryagin set difference operator. It is easy to see, that if Assumption~\ref{as:1} is satisfied for some $\gamma>0$, then generating training data with features $\Phi_{\gamma}(\cdot)|_{x_M}$ and labels $\Phi_{\gamma}(\cdot)|_{x_{M+1},\hdots,x_N}$, which yield the predictions $\tilde{x}\in \mathcal X$. This, in turn, implies that the Neural Horizon MPC formulation \eqref{eq:neural_MPC} with $\mathcal X_k = \tilde{\mathcal X}_k = \mathcal X$ for every $k\in [0,\hdots,N]$ will remain feasible for every admissible initial state of $\Phi_\gamma(\cdot)$.

Furthermore, the formulation \eqref{eq:neural_MPC} with a Neural Horizon constructed from such a $\Phi_\gamma(\cdot)$, can be viewed as a simplified case of the MPC formulated in \cite{Baethge2016}. If Assumption~\ref{as:1} holds, the error bound over the Neural Horizon is constant and independent of the state, which can be used directly, without the construction of a tube-based MPC over the second horizon. Besides, as we do not need a projection at $x_M$ in \eqref{eq:neural_MPC} to interface between the horizons, we satisfy all assumptions and requirements of \cite{Baethge2016}. Thus, Corollary~\ref{c:1} follows from Proposition 3.2 of \cite{Baethge2016}:
\begin{corollary} \label{c:1}
    If for $\gamma>0$ a neural network trained on $\Phi_\gamma(\cdot)$ satisfies Assumption~\ref{as:1}, then for the state constraints $\mathcal X_k=\mathcal X$, $k\in[0,\hdots,M]$ and $\tilde{\mathcal X}_k=\mathcal X$, $k\in[M+1,\hdots,N]$, the Neural Horizon MPC \eqref{eq:neural_MPC} is recursively feasible for all $t \geq t_0$.
\end{corollary}
Note, that under Assumption~\ref{as:1} the constraints $\tilde{x}_k\in \mathcal{X}$ for $k\in[M+1,\hdots,N]$ in \eqref{eq:neural_MPC} are satisfied by construction, and thus can be lifted from the formulation.



\section{Simulation Results}
\label{sec:simresults}

\begin{table}[!h]
    \caption{State and input constraints}
    \begin{center}
        \begin{tabular}{r|l@{\ }l@{\ }l}
            \toprule
            \textbf{\textit{Variable}} & \textbf{\textit{Lower bound}} & \textbf{\textit{Upper bound}} & \textbf{\text{Description}} \\
            \midrule
            $x_{\mathrm{cart}}\in \mathcal X^1$ & $-2~m$ & $2~m$ & Length of the rail  \\
            \trowgray $\theta\in \mathcal X^2$ & $-6\pi$ & $6\pi$ & Angle of the pendulum  \\
            $v\in\mathcal X^3$ & $-10~ms^{-1}$ & $10~ms^{-1}$ & Cart velocity   \\
            \trowgray $\omega\in\mathcal X^4$ & $-10~s^{-1}$ & $10~s^{-1}$ & Angular velocity  \\
            $F\in\mathcal U\phantom{_k}$ & $-80~Nm$ & $80~Nm$ & Force on the cart  \\
            \bottomrule
        \end{tabular}
    \label{tab:constr}
    \end{center}
\end{table}

\begin{table}[!h]
    \caption{Model Parameters}
    \begin{center}
        \begin{tabular}{c|r l l}
            \toprule
            \textbf{\textit{Parameter}} & \textbf{\textit{Value}} & \textbf{\textit{Unit}} & \textbf{\text{Description}} \\
            \midrule
            $M$ & 1 & $[kg]$ & Mass of the cart  \\
            \trowgray $m$ & 0.1 & $[kg]$ & Mass of the pendulum  \\
            $l$ & 0.8 & $[m]$ & Length of the pendulum  \\
            \trowgray $g$ & 9.81 & $[m/s^2]$ & Gravitational acceleration  \\
            \bottomrule
        \end{tabular}
    \label{tab:para}
    \end{center}
\end{table}

\begin{figure}[!h
]
    \centerline{
        \input{pendulum_figure}
    }
    \caption{Inverted pendulum system. Input force $F$ moves the cart. The origin is set to the angle $\theta$ in the upright position and the cart $x_{\mathrm{cart}}$ in the middle of the rail.}\label{fig:pend}
\end{figure}
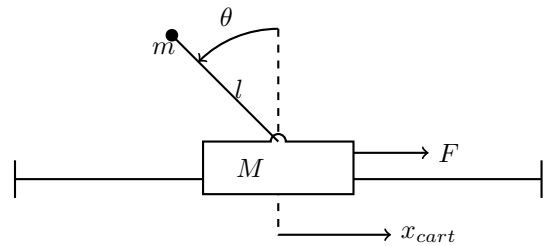


We consider a model of an inverted pendulum on a cart (cf. Fig.~\ref{fig:pend}), which runs on a rail of a fixed length, given by the positional constraint in TABLE~\ref{tab:constr}. In continuous time this can be modelled by

\begin{align*}
     &\dot{x}_{\mathrm{cart}} = v, &&\quad \dot{v} = \frac{\mu_1(\theta,\omega)\cos(\theta) + F + g m \cos(\theta) \sin(\theta) }{\mu_2(\theta)}\\
     &\dot{\theta} = \omega, &&\quad \dot{\omega} = \frac{\mu_1(\theta,\omega) + F \cos(\theta)}{l \mu_2(\theta)},
\end{align*}
with $\mu_1(\theta,\omega)\!=\!-l m \sin(\theta) \omega^2$ and $\mu_2(\theta)\!=\!M\!+\!m(1\!-\!\cos^2(\theta))$. Here, $x_{\mathrm{cart}}$ represents the cart position, $v$ -- its velocity, $\theta$ denotes the angle of the pendulum and $\omega$ -- the angular velocity. Remaining model parameters are shown in TABLE~\ref{tab:para}. Accordingly, we define the states as $x = [x_{\mathrm{cart}},\theta,v,\omega]^T$ and the input is $u = F$. The $x_{\mathrm{cart}}=0$ is set as the middle of the rail and the angle of the pendulum $\theta=0$ is chosen in the upright position. This model is used for the considered controllers as well as the simulation of the real system. The model is discretized with $\Delta t=20\mathrm{ms}$ and numerically integrated via a Runge-Kutta method of order $4$.

\subsection{Baseline MPC Horizon}
\label{subsec:simnominal}

\begin{figure}[!ht]
    \centering
    \includegraphics[width=\columnwidth]{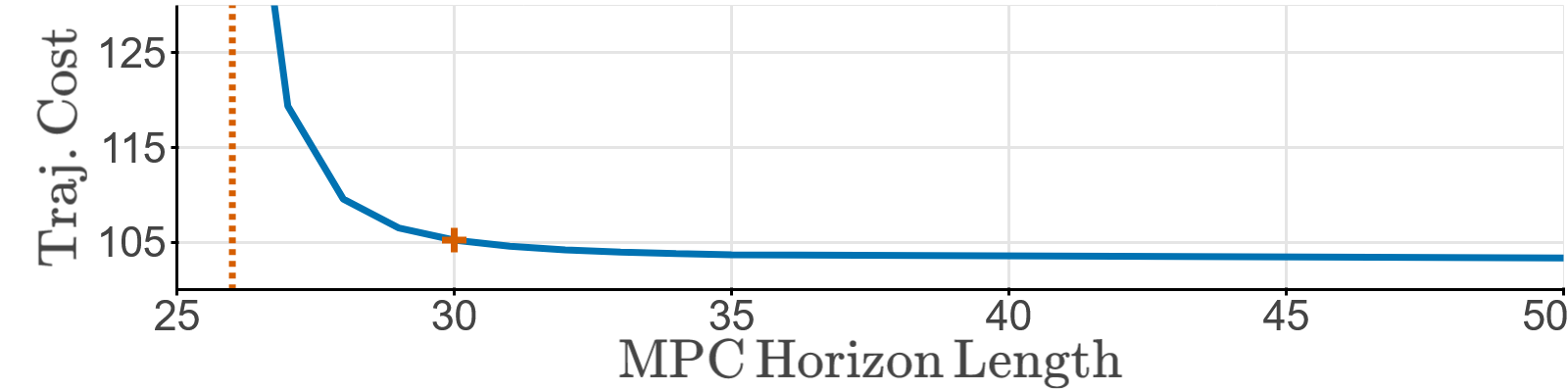}
    \caption{Comparison of trajectory costs for the MPC controllers with respect to the prediction horizon lenghts.}
    \label{fig:mpc-horlen}
\end{figure}

For the qualitative comparison of controller performance, we require a MPC that can stabilize the system for a reasonably short horizon. To find this MPC we conducted several closed-loop simulations. Fig.~\ref{fig:mpc-horlen} shows the relation between the MPC horizon length and the overall cost of the $5\mathrm{s}$ simulation window, starting from the same initial conditions. The horizon length of $N=30$ steps was chosen, as the controllers with longer horizons only provide a limited improvement in quality, while taking significantly longer to solve each underlying OCP step. Shorter horizon lengths show significant degradation of the quality, below $27$ steps the MPCs fail to stabilize the system.

\subsection{Controllers}
\label{subsec:controllers}
To evaluate the quality and computational improvements of the proposed approaches, we compare the following controller designs:
\begin{itemize}
        \item \textbf{Baseline MPC} is based on the OCP formulation \eqref{eq:mpc}. We use a quadratic cost function\vspace{-.5em}
        \[L\left(\{\!x_k\!\}_0^{N-1}\!,\{\!u_k\!\}_0^{N-1}\right)\!=\!\textstyle\sum_{k=0}^{N-1} \lVert x_k \rVert^{2}_{Q} +\lVert u_k \rVert^{2}_{R},\]
        and  $V(x_N)\!=\!\lVert x_N \rVert^{2}_{Q}$, where $\lVert \xi_k \rVert^{2}_{W}=\xi_k^T W \xi_k$ denotes a weighted $\ell^2$-norm. $Q$ and $R$ represent weight matrices. This controller is also used to generate data for the training of every neural network.
         
        \item \textbf{Short horizon MPC} is equivalent to Baseline MPC, but with a horizon length of $M=8$.
        
        \item \textbf{Imitation Controller} approximates the controller itself using a neural network, similar to the approach in~\cite{Borrelli2019}. This network is implemented and trained via supervised learning to approximate the MPC mapping~\eqref{eq:MPCmapping} from the current sample point $x(t_i)$ to the optimal first input $u_0^*(t_i)$, $\tilde f_{\mathrm{imit}}: \mathbb{R}^{n_x} \rightarrow \mathbb{R}^{n_u}$.
        
        \item \textbf{Neural Horizon MPC} implements the Neural Horizon MPC \eqref{eq:neural_MPC} with the state sequence approximation. We use the quadratic cost for the second horizon $\tilde L \left( \tilde x_k \right) = \lVert \tilde x_k \rVert^{2}_{Q},\ \forall k \in [M\!+\!1,\hdots,N]$, where the weight matrix $Q$ is the same as in the Baseline MPC. Additionally, we include a variant of this controller without state constraints $\tilde{x}\in\tilde{\mathcal{X}}_k$ in \eqref{eq:neural_MPC}.
        
        \item \textbf{Cost-Estimation MPC} is the variant with direct cost estimation \eqref{eq:nn_cost_model}, described in Section~\ref{subsec:nhconstruction}.
\end{itemize}

\subsection{Neural Network Implementation}
\label{subsec:NNImpl}
All approximation-based controllers introduced in Section~\ref{subsec:controllers} employ neural networks with $3$ hidden layers, $32$ neurons, and \emph{tanh} activation functions (these and other hyperparameters were determined via grid search).
For consistency, a single dataset was generated to train all three NNs, contraining $25000$ datapoints of the form $\{x^i,\Phi_\gamma(x^i)\}_{i=1}^{25000}$ with $\gamma=0.25$ (cf. Section~\ref{subsec:stab}), i.e. we have recorded the predicted optimal states and inputs of the OCP~\eqref{eq:mpc} with the prediction horizon $N$. This means our networks were trained purely on optimal open-loop sequences as they are calculated at a given sampling point. The initial states $x^i$ were chosen randomly from a subset $0.75\mathcal X^1\!\times\!0.25\mathcal X^2\!\times\!0.25\mathcal X^3\!\times\!0.25\mathcal X^4$ (cf. TABLE~\ref{tab:constr}) $10~\mathrm{\%}$ of the time, or derived from the previous datapoint prediction $x^{i+1}:=\Phi_\gamma(x^i)|_{x_1}$ $90~\mathrm{\%}$ of the time.
For the Neural Horizon MPC with the horizons $(M,N)=(8,30)$, as shown in Section~\ref{subsec:stab}, we set $\Phi_\gamma(x^i)|_{x_M}$ as features and $\Phi_\gamma(x^i)|_{x_{M+1},\hdots,x_N}$ as labels. For the Cost Estimation version, the same features were chosen, but the labels were constructed from the corresponding variables according to \eqref{eq:nn_cost_model}. The Imitation Controller was trained on $x^i$ and $\Phi_\gamma(x^i)|_{u_0}$ as features and labels correspondingly. To counteract the stochasticity of the training process, we have trained each of the networks $10$ times and selected $5$, that yielded the lowest closed-loop trajectory costs (or retained feasibility the longest).
All NNs were trained for $2500$ epochs and a different validation dataset of $5000$ points, generated in the same manner, was used to calculate the $R^2$-score shown in TABLE~\ref{tab:r2}.
It took about $600$ seconds to fully train each Neural Network, though given that this step is performed offline, it should not matter for comparing the closed-loop performance of the controllers.

\begin{table}[!htbp]
    \caption{$R^2$-scores for implemented neural networks}
    \begin{center}
        \begin{tabular}{ll|rc}
            \toprule
            \textbf{\textit{Neural Network}} & \textbf{\textit{Handle}} &
            \textbf{\textit{Outputs}} & $\mathbf{R^2}$\textbf{\textit{-score}} \\ \midrule
             State Approximation & $\tilde f_{s}(\cdot)$ & $88$ & $[0.992,0.994]$  \\
            \trowgray Cost Approximation & $\tilde f_{c}(\cdot)$ & $2$ & $[0.988,0.989]$ \\
            Imitation & $\tilde f_{\mathrm{imit}}(\cdot)$ & $1$ & $[0.951,0.960]$  \\
            \bottomrule
        \end{tabular}
    \label{tab:r2}
    \end{center}\vspace{-4 ex}
\end{table}

We simulated the closed-loop of an inverted pendulum controlled by all approaches presented in Section \ref{subsec:controllers}.
The horizon length $N=30$ was chosen for the Baseline MPC (cf. Fig.~\ref{fig:mpc-horlen}), and to produce training data for $\Phi_\gamma(\cdot)$ with $\gamma=0.25$ (cf. Section~\ref{subsec:stab}). 
To compare computational efficiency, we used the average wall time of a single controller timestep, simulated using CasADi \cite{Andersson2019} framework with IPOPT nonlinear solver. The networks were trained using PyTorch and converted to the CasADi symbolic formulation as well. The reference point was set to $0$ for all states. The bounds for the OCP are provided in TABLE \ref{tab:constr}, which for simplicity are also set as the terminal constraints.\\
All simulations were run on a desktop PC with Ubuntu-based OS, an AMD Ryzen $9$ $5950$X $16$-Core CPU, $64~\mathrm{GB}$ RAM, and the neural networks were trained using Nvidia GeForce RTX $3090$ GPU.

\subsection{Closed-loop Simulations}
\label{subsec:simneural}

\begin{figure}[!b]
    \includegraphics[width=0.95\columnwidth]{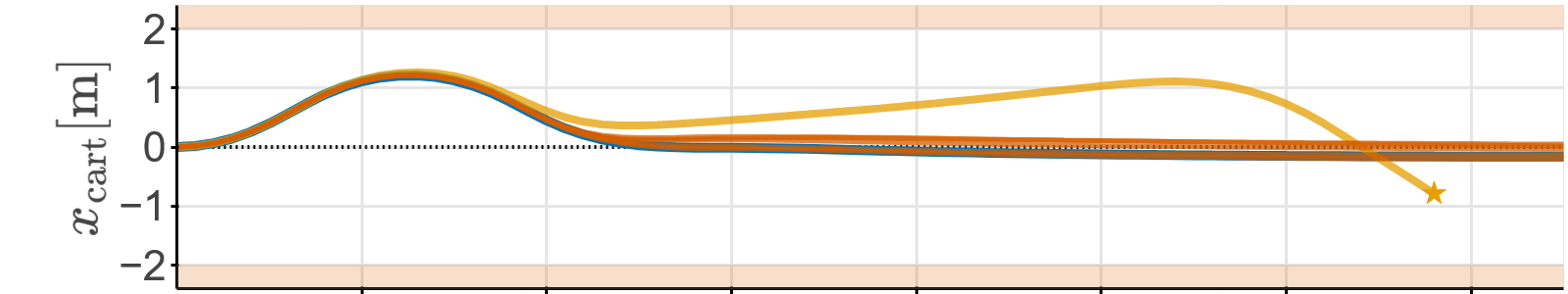}
    \includegraphics[width=0.95\columnwidth]{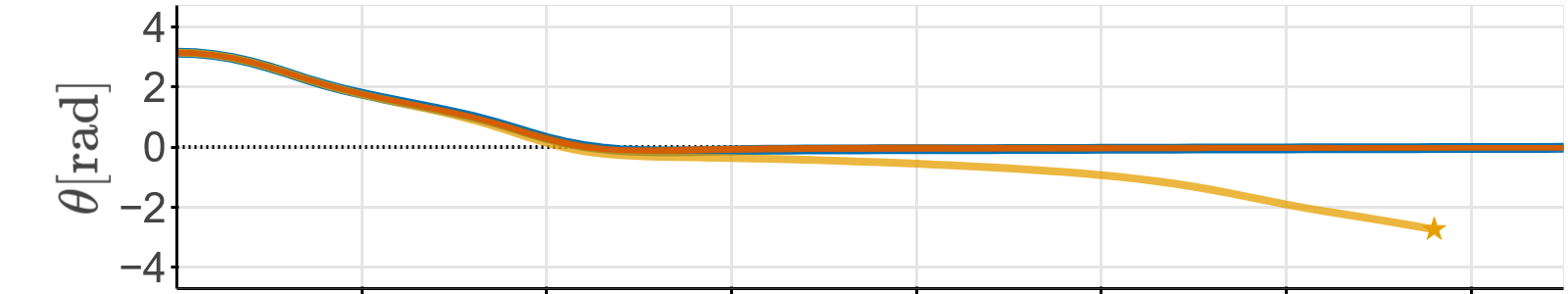}
    \includegraphics[width=0.95\columnwidth]{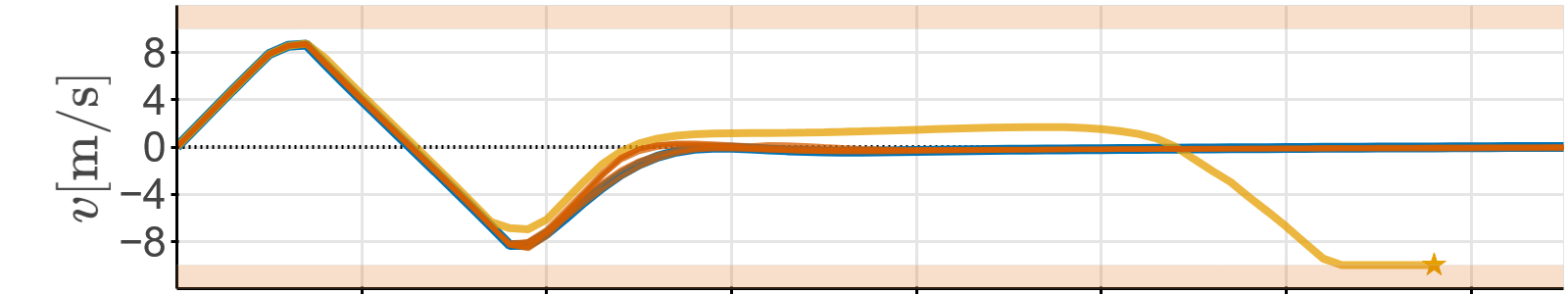}
    \includegraphics[width=0.95\columnwidth]{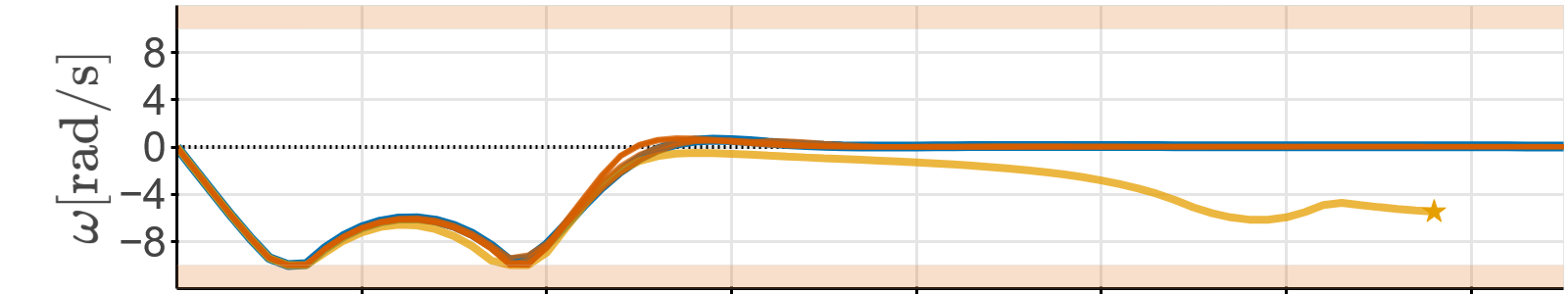}
    \includegraphics[width=0.95\columnwidth]{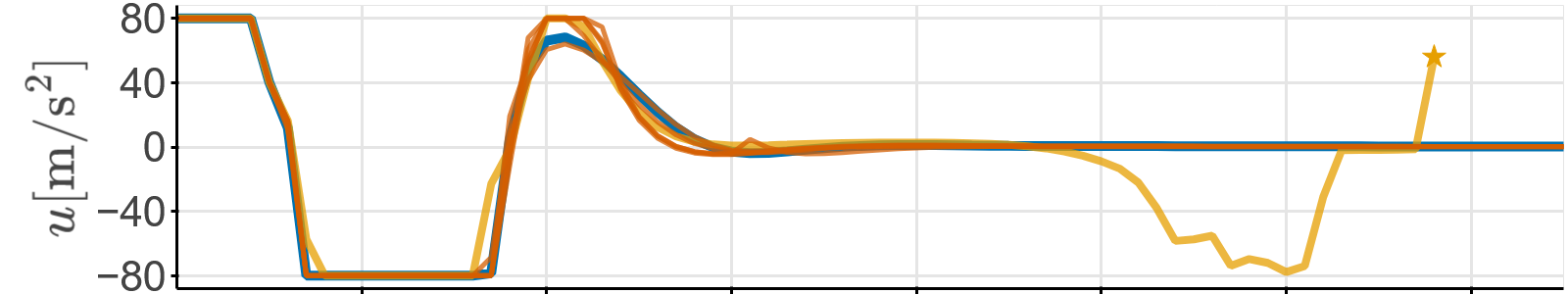}
    \includegraphics[width=0.95\columnwidth]{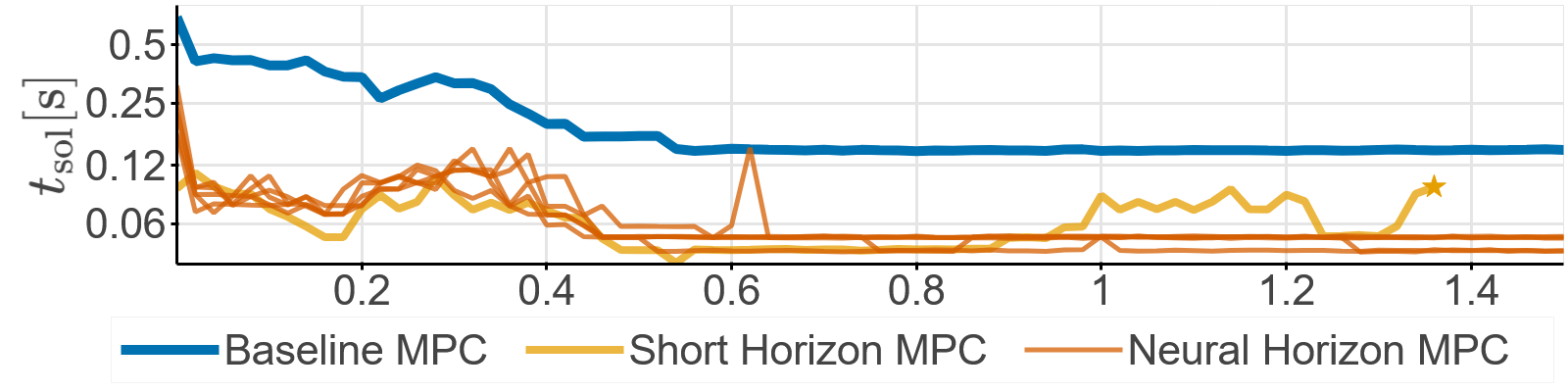}
    \caption{Upswing of the inverted pendulum on a cart. Trajectories are plotted along horizontal axis (in seconds). Star denotes the point when Short horizon MPC became infeasible.}
    \label{fig:nh-mpc-res1}
\end{figure}

We simulate the upswing of the pendulum with all controllers for $5\mathrm{s}$ of simulation time. The comparison between the Short Horizon MPC, Baseline MPC, and Neural Horizon MPC ($5$ trained NNs) with explicit state bounds are shown in Fig.~\ref{fig:nh-mpc-res1}. Here, the Baseline MPC and all Neural Horizon MPCs manage to reach the reference and stabilize the system. The Short Horizon MPC fails to achieve the goal and becomes infeasible at $1.38~\mathrm{s}$ of simulation time. The solver wall time of the Neural Horizon MPCs is lower than one of the Baseline MPC at every iteration, and is lower than the solver wall time of the Short Horizon on average (cf. also TABLE~\ref{tab:perfromancecompare}).

\begin{figure}[!b]
    \includegraphics[width=0.95\columnwidth]{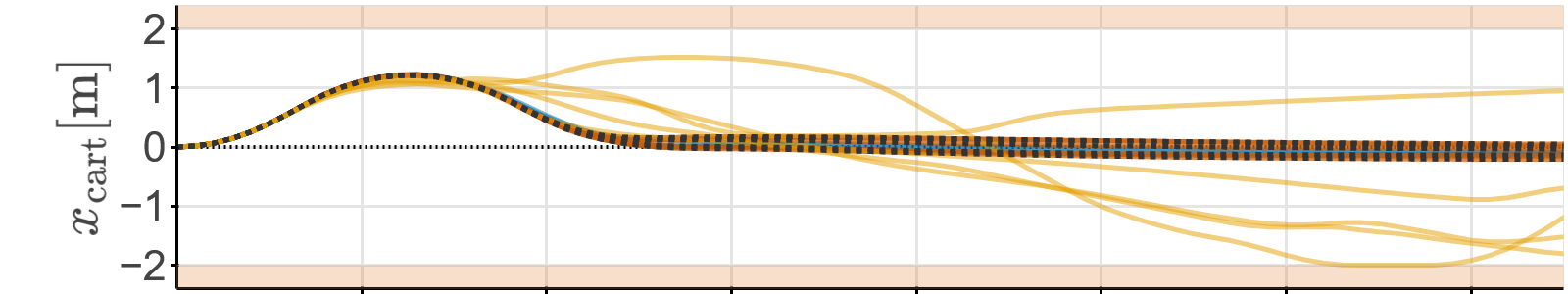}
    \includegraphics[width=0.95\columnwidth]{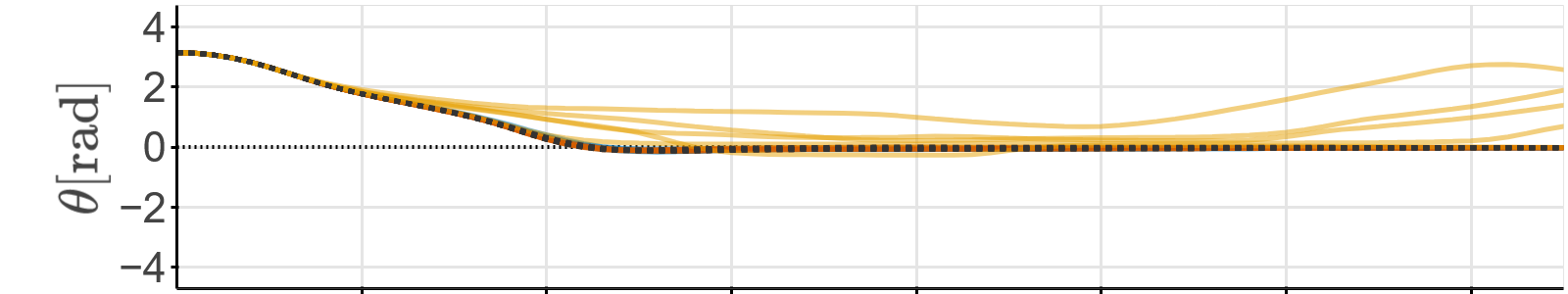}
    \includegraphics[width=0.95\columnwidth]{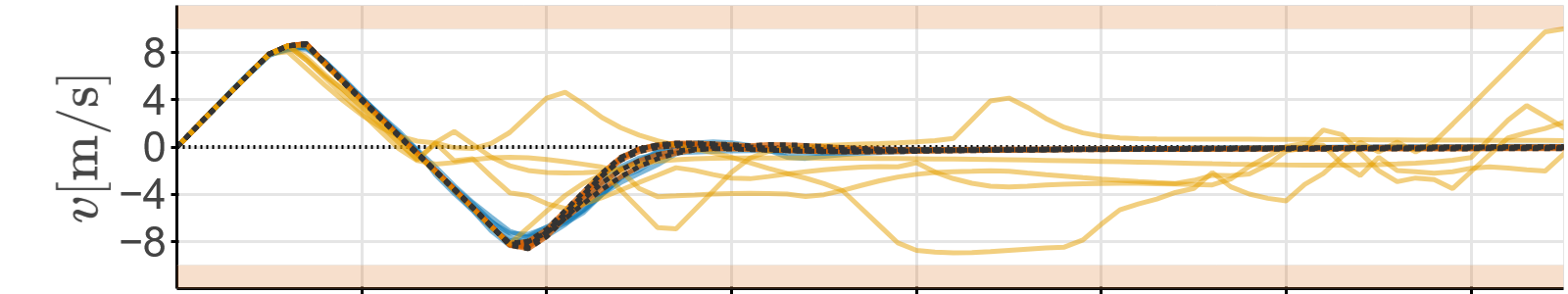}
    \includegraphics[width=0.95\columnwidth]{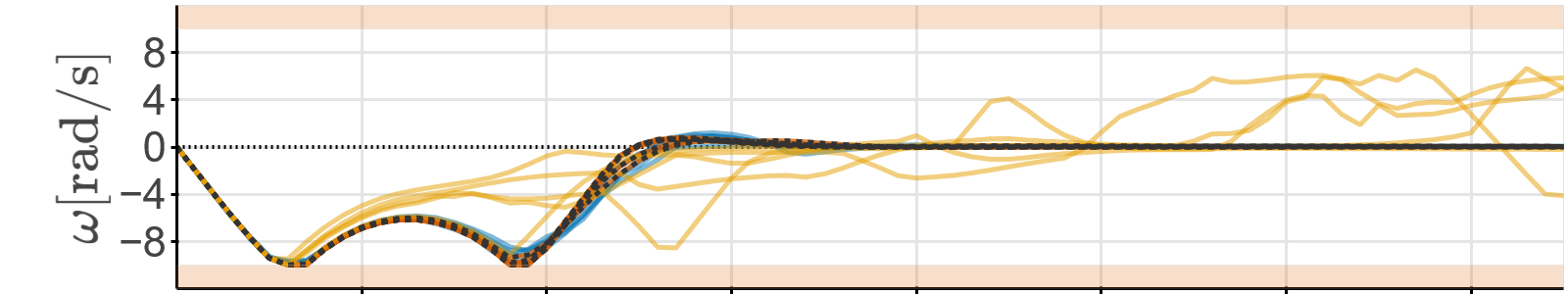}
    \includegraphics[width=0.95\columnwidth]{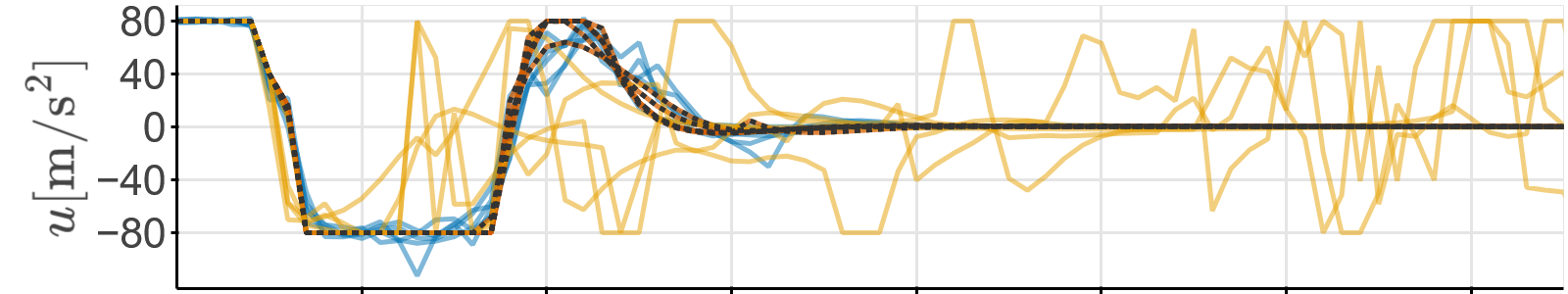}
    \includegraphics[width=0.95\columnwidth]{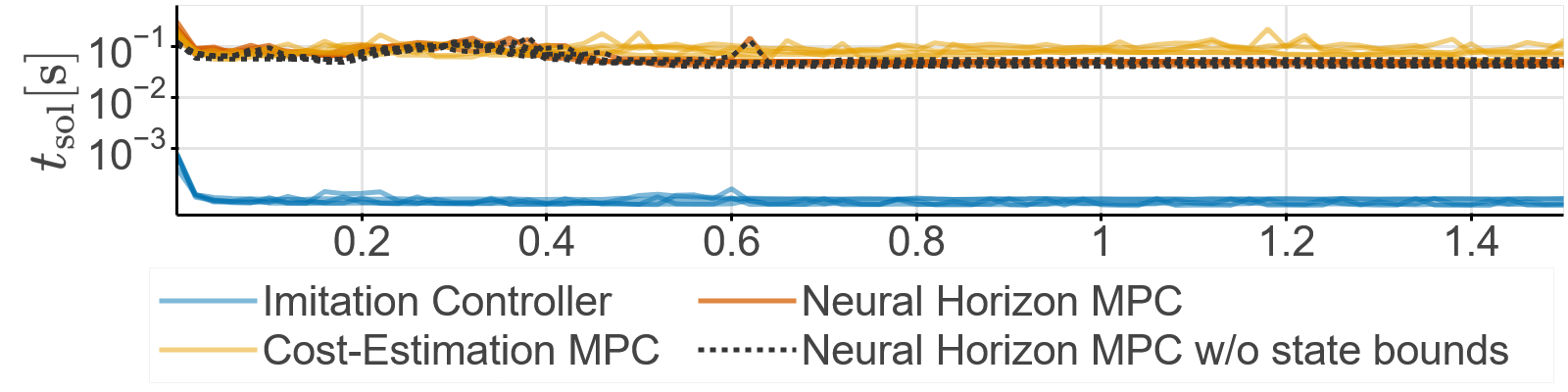}
    \caption{Upswing of the inverted pendulum on a cart. Trajectories are plotted along horizontal axis (in seconds).}
    \label{fig:nh-mpc-res2}
\end{figure}

A comparison between all approximation-based controllers is shown in Fig.~\ref{fig:nh-mpc-res2}. Here, the Cost-Estimation MPCs cannot stabilize the system and in some cases become infeasible (for $6$ out of $10$ trained NNs). The Neural Horizon MPCs without state bounds are slightly faster (cf. TABLE~\ref{tab:perfromancecompare}), but produce identical trajectories to the version with explicit state bounds. The Imitation Controllers have also stabilized the reference, though most of them have violated input constraints. Given that these controllers are static mappings, their computation time was also much lower than that of other considered controllers.

\begin{table}[!b]
    \caption{Closed-loop metrics of controllers}
    \begin{center}
        \begin{tabular}{l@{\ }|l@{\ }l}
            \toprule
            \textbf{{Controller}} &\textbf{Iter. Wall Time [$\mathrm{ms}$]} & \textbf{{Trajectory Cost}} \\
            \midrule
            Baseline& \eqmakebox[pm][r]{$160\;\pm\;$}$65$ & \eqmakebox[bnd][l]{}$105.23$  \\
            \trowgray Short Horizon \textbf{*} &\eqmakebox[pm][r]{$63\;\pm\;$}$18$ & \eqmakebox[bnd][l]{}$203.67$ \\
            Cost-Estimation \textbf{*} & \eqmakebox[pm][r]{$89\;\pm\;$}$26$ & \eqmakebox[bnd][l]{$[385.11,$}$1801.98]$  \\
            \trowgray Neural Horizon & \eqmakebox[pm][r]{$\mathbf{53}\;\pm\;$}$\mathbf{17}$& \eqmakebox[bnd][l]{$[\mathbf{103.51},$}$107.22]$  \\
            Neural Horizon w/o s. b. & \eqmakebox[pm][r]{$\mathbf{49}\;\pm\;$}$\mathbf{12}$ & \eqmakebox[bnd][l]{$[\mathbf{103.51},$}$107.22]$  \\
            \trowgray Imitation Controller & \eqmakebox[pm][r]{$\mathbf{0.09}\;\pm\;$}$\mathbf{0.04}$ & \eqmakebox[bnd][l]{$[104.39,$}$106.08]$  \\
            \bottomrule
        \end{tabular}
    \label{tab:perfromancecompare}
    \end{center}
\end{table}


\section{Performance Discussion}
\label{sec:discussion}
As illustrated in Fig.~\ref{fig:mpc-horlen}, the controller performance falls sharply for a horizon length lower than $30$. However, we showed that the Neural Horizon recovers the missing information and can approximate it well enough to reduce the first horizon down to 8 steps. Despite the fact that mathematically the Cost-Estimation MPC version is very similar to the Neural Horizon MPC without explicit state bounds, we observe a much worse performance even for comparably trained neural networks (cf. TABLE~\ref{tab:r2}). A possible reason is the complex relation between the state $x_M$ and the remaining cost terms, that the Neural Horizon formulation regularizes by aggregating separate state predictions. However, a follow-up investigation is required.

Among the stabilizing controllers, the Imitation Controller has the fastest computations, as shown in TABLE~\ref{tab:perfromancecompare} since it does not perform any optimization and simply queries the NN approximation for the next control input. Employing such a controller in a real control system should thus be done with caution, as it has no way of strictly adhering to the constraints. Though state constraints were not violated in our simulations, this cannot be guaranteed in the general case. It has violated the input constraints, which could be avoided e.g. by projections \cite{zieger2021one}, but it could lead to further quality degradation.
In contrast, the Neural Horizon MPC with state sequence approximation could stabilize the system, while guaranteeing that no state or input constraints are violated in the closed-loop. This is achieved by construction, as the optimal predicted trajectory is still bounded in the first horizon. The Neural Horizon, on the other hand, reduces the computational burden of the algorithm to be on par with the Short Horizon MPC, which shows a slightly higher computation time in simulations as it fails to converge. Though in our simulations explicit constraints over the Neural Horizon did not play a role, further research is needed for cases when processes operate close to state boundaries.


\section{Conclusion and Outlook}
\label{sec:conclusion&outlook}

We have demonstrated a method to successfully reduce the horizon length of the MPC algorithm by employing neural networks to construct an approximation of the horizon tail. We have further shown how these networks can be utilized inside the optimization problems, despite their adverse gradient properties.
Though in this work we have concentrated on the most direct comparison between the Baseline MPC and the proposed Neural Horizon MPC, there are steps to make this approach even more practical. As the training data is generated offline, we can use a controller with a much longer time horizon, in essence approximating an infinite-horizon MPC formulation. On the other hand, we can use any source of control actions for training, e.g. emulating a human driver to derive realistic control actions while retaining constraint satisfaction from the first horizon of the MPC.

Though our simulation results indicate that explicit constraints in the Neural Horizon are not required for maintaining close-to-optimal performance, further research is needed to confidently state that.
A further investigation of the poor performance of the Cost-Estimation MPC is also needed to determine possible conditions for this approach to perform on par with the Neural Horizon variant.\vspace{-0.5em}


\section*{ACKNOWLEDGMENT}
The authors acknowledge funding of the KI-Embedded project of the German Federal Ministry of Economic Affairs and Climate Action (BMWK).

\bibliographystyle{IEEEtran}
\bibliography{neural_horizon_literature,IEEEabrv}
\end{document}

%% file: pendulum_figure.tex
\begin{tikzpicture} [thick]

\newcommand{\ang}{-45}

\draw [black!80!black, -] (-3.5,0.5) -- (3.5,0.5);
\draw [black!80!black, -] (-3.5,0.25) -- (-3.5,0.75);
\draw [black!80!black, -] (3.5,0.25) -- (3.5,0.75);

\draw[dashed] (0,2.5) -- (0,-0.25) node (vert) [very near start]{};

\draw[black!80!black, ->] (0,-0.24) -- (1.5,-0.24) node [right] {$x_{cart}$};

\begin{scope} [draw = black,fill = white!20,dot/.style = {white, radius = .025}]

\filldraw (-1,1) -- coordinate [pos = .5] (F) (-1,.3) -- node [above = 0.35cm, left = 0.05cm] {$M$} (1,.3) -- (1,1)  coordinate (X) -- (.1,1) arc (0:180:.1) -- (-1.014,1);

\draw [black!80!black,fill = black,rotate around = {-\ang:(0,1)}] (0,1) -- node (pend)  [midway] {$ $} node [midway,right] {$l$} node [very near end, left] {$m$} +(0,2) circle (0.07) coordinate [pos = .5] (T);

\draw [black,->] (1,0.85) -- (2,0.85) node [black,right] {$F$};

\fill [dot] (0,1.52) circle;

\coordinate  (pendbottom) at (0,1);

\pic [draw=black, text=black, ->, "$\theta$", angle eccentricity=1.2,angle radius=1.5cm] {angle = vert--pendbottom--pend};

\end{scope}

\end{tikzpicture}